\documentclass[conference]{IEEEtran}
\IEEEoverridecommandlockouts
\usepackage{cite}
\usepackage{amsmath,amssymb,amsfonts}
\usepackage{algorithmic}
\usepackage{graphicx}
\usepackage{textcomp}
\usepackage{xcolor}
\usepackage{color}
\usepackage{multirow}
\usepackage{subfigure}
\def\BibTeX{{\rm B\kern-.05em{\sc i\kern-.025em b}\kern-.08em
    T\kern-.1667em\lower.7ex\hbox{E}\kern-.125emX}}
\usepackage{caption}
\begin{document}

\title{A parallelizable variant of HCA*
\vspace{-0.5cm}
}
\author{
\IEEEauthorblockN{Sreenivasan Ganti}
\IEEEauthorblockA{\textit{Accenture Labs} \\
\textit{Bengaluru, India}\\
sreenivasan.ganti@accenture.com}\\

\IEEEauthorblockN{Shravan Mohan\textsuperscript{†}}
\IEEEauthorblockA{\textit{Accenture Labs} \\
\textit{Bengaluru, India}\\
shravan.rammohan@gmail.com}
\thanks{\textsuperscript{†}Shravan contributed to this work when he was part of Accenture Labs, Bengaluru}

\and

\IEEEauthorblockN{Visnu Srinivasan}
\IEEEauthorblockA{\textit{Accenture Labs} \\
\textit{Bengaluru, India}\\
visnu.srinivasan@accenture.com}\\

\IEEEauthorblockN{Milind Savagaonkar}
\IEEEauthorblockA{\textit{Accenture Labs} \\
\textit{Bengaluru, India}\\
milind.savagaonkar@accenture.com}

\and

\IEEEauthorblockN{Pallavi Ramicetty\textsuperscript{*}}
\IEEEauthorblockA{\textit{Accenture Labs} \\
\textit{Bengaluru, India}\\
pallavi.ramicetty@gmail.com}
\thanks{\textsuperscript{*}Pallavi contributed to this work when she was part of Accenture Labs, Bengaluru}\\
\IEEEauthorblockN{Shubhashis Sengupta}
\IEEEauthorblockA{\textit{Accenture Labs} \\
\textit{Bengaluru, India}\\
shubhashis.sengupta@accenture.com}
}


\maketitle

\begin{abstract}
This paper presents a parallelizable variant of the well-known Hierarchical Cooperative A* algorithm (HCA*) for the multi-agent path finding (MAPF) problem. In this variant, all agents initially find their shortest paths disregarding the presence of others. This is done using A*. Then an intersection-graph (IG) is constructed; each agent is a node and two nodes have an edge between them if the  paths of corresponding agents collide. Thereafter, an independent set is extracted with the aid of an approximation algorithm for the maximum independent set problem. The paths for the agents belonging to independent set are fixed. The rest of agents now again find their shortest paths, this time ensuring no collision with the prior agents. Space-time A*, which is a crucial component of HCA*, is used here. These iterations continue until no agents are left. Since the tasks of finding shortest paths for the agents in any iteration are independent of each other, the proposed algorithm can be parallelized to a large extent. In addition to this, the task of determining the IG can also be done in parallel by dividing the map into sections and with each agent focusing on a particular section. The parallelism does come at a cost of communication between the agents and the server. This is accounted for in the simulations. As an added advantage, the user need not make a choice for the priority order. It is observed, empirically, that the proposed algorithm outperforms HCA* in terms of the computation time and the cost value in many cases. Simulations are provided for corroboration. 
\end{abstract}

\vspace{-0.2cm}
\section{Introduction}
\noindent \textbf{MAPF, its variants and solution methodologies}: The task of multi-agent path-finding (MAPF) is central to the field of robotics \cite{lavalle2006planning}. With the advent of robotic warehouses which operate hundreds of robots, MAPF has gained much attraction in the recent years \cite{ma2017ai}. In such scenarios the computational speed of path finding, while maintaining low costs, is important. MAPF has several variants, all of which are important. For example, the MAPF with kinematic constraints (MAPF-KC) \cite{honig2016multi} requires one to consider finite acceleration/deceleration of robots, whereas the MAPF combined with target assignment (TAPF) \cite{ma2016optimal} requires one to determine the optimal assignment of tasks along with path finding. However, the focus of this paper would be the conventional MAPF problem (this is described in detail later). The MAPF problem is NP-hard \cite{yu2013structure}, essentially due to the requirement of collision avoidance. There are exact methods such as Conflicts Based Search (CBS) \cite{sharon2015conflict}, and there are heurisitics such as (Hierarchical Cooperative A*) HCA* \cite{silver2005cooperative}. The exact methods are typically computationally inefficient, while heuristic methods, although fast, typically yield sub-optimal results. This paper focuses on the popular heuristic called HCA*, and proposes a parallelizable variant of it. \\
\begin{figure}{}
    \centering
    \includegraphics[width=3.5in]{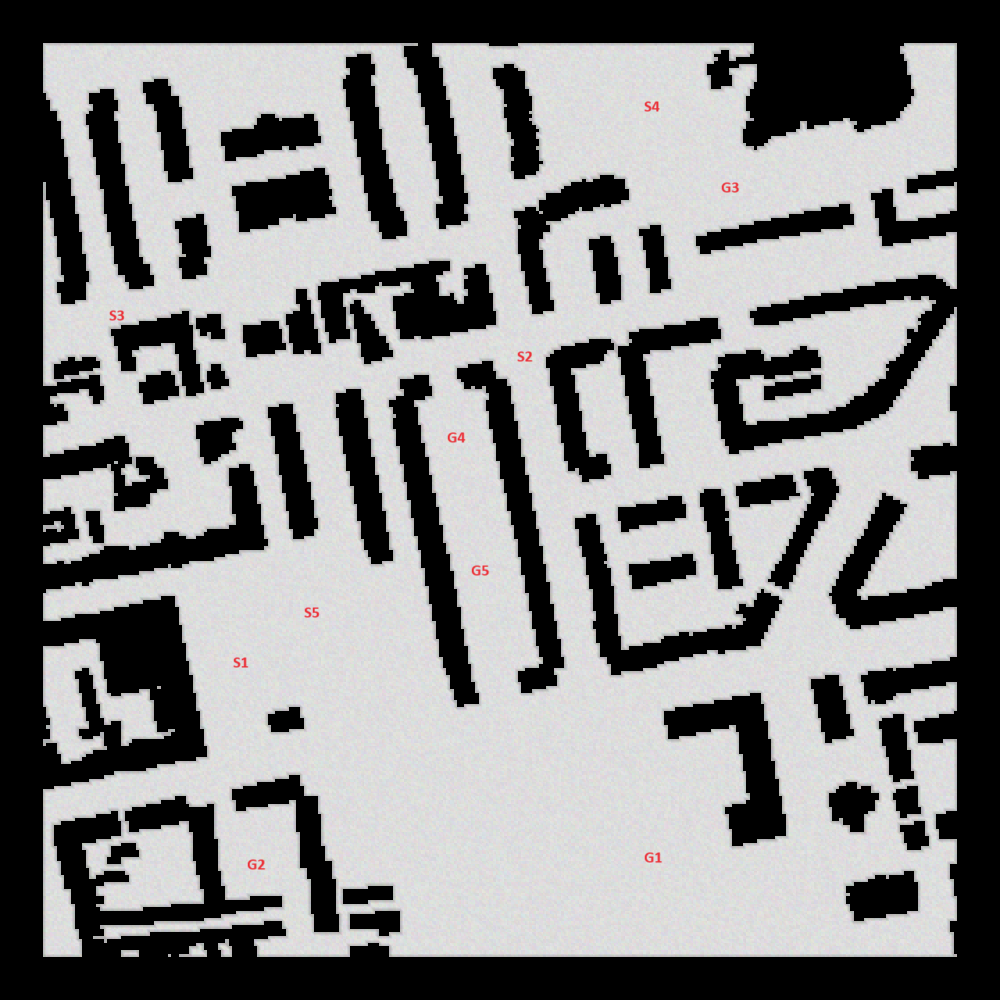}
    \caption{\small A typical problem configuration for the MAPF problem. This image shows the Berlin Map \cite{stern2019mapf} with source and goal locations for 5 agents marked by $S_1, S_2, S_3, S_4, S_5$ and $G_1, G_2, G_3, G_4, G_5$, respectively. The task is to find non-colliding paths for the 5 agents such that the sum of their path lengths (including waiting times) is minimum. This map is also used for benchmarking.}
    \label{fig:berlin_map}
    \vspace{-0.7cm}
\end{figure}
\noindent \textbf{Prior art}: The idea of parallelism in computation and in path finding has been pursued earlier. For example, the algorithm called GA*, proposed in \cite{zhou2015massively}, essentially uses multiple priority queues to process open nodes list in parallel by the GPU cores, as required by the A* algorithm. This can potentially provide a speedup of up to 45x. It can also be noted that GA* can be used in conjunction with the method presented in this paper. Similarly, the idea of detecting subsets of agents whose paths can be found independently of each other has been explored in the past. The idea proposed by \cite{standley2012independence} is one such. The groups themselves are formed as the algorithm proceeds. Initially, each agent is assigned its own group and shortest paths are calculated. In case of a conflict, the groups involved in the conflict are combined and non-colliding paths for the agents in the combined group are found again. This is repeated until non-colliding paths are found for every agent. On the other hand, the authors in \cite{mogali2020template} propose a Lagrangian Relax-and-Cut (LRC) procedure to determine a lower bound for the MAPF problem, which is then used for node evaluation in CBS. Within the LRC procedure, the authors determine the maximum independent set from an intersection graph as an initial step, but then the paths for the rest of the agents are determined in a specific order. In particular, the most constrained agent gets the highest priority. In contrast to the aforementioned references, and to the best of the authors' knowledge, a focused effort on parallelizing HCA* has not been undertaken. 
\begin{figure}[t]
\centering
\fbox{\begin{minipage}{3.3in}
\begin{enumerate}
    \item Initialize the \textit{set of start \& goal locations} $S$ and $G$ to the empty set.
    \item Let $O$ be the \textit{set of static obstacles} in the map.
    \item Let the \textit{free space} set $F$ be initialized to $O^c$.
    \item Let $\sigma$ be a \textit{random permutation}. 
    \item {For $i = 1$ to $N$:\\
    \hspace*{0.5cm} Choose the agent $\sigma(i)$.\\
    \hspace*{0.5cm} Choose \textit{distinct} $s$ and $g$ randomly from $F$.\\
    \hspace*{0.5cm} While there is \textit{no path} $P$ between $s$ and $g$:\\
        \hspace*{1cm} Choose distinct $s$ and $g$ randomly from $F$.\\
    \hspace*{0.5cm} Update $S = S\bigcup \{s\}$ and $G = G\bigcup \{g\}$.\\
    \hspace*{0.5cm} Update $O = O\bigcup \{s,g\}$.\\
    \hspace*{0.5cm} Update $F = F  - P$.
    \item Return $S$ and $G$.
    }
\end{enumerate}
\end{minipage}}
\caption{\small A pseudocode of the instance generation routine shown above generates a problem instance ensuring that prioritized path planning yields a solution under any priority order.}
\label{instanvce_gen}
\vspace{-0.5cm}
\end{figure}

\noindent \textbf{Windowed HCA*}: HCA* is an incomplete heuristic, which was the primary motivation for the advent of windowed HCA* (WHCA*) \cite{silver2005cooperative}. In WHCA*, one starts with a priority order and finds paths using HCA* for agents, such that those are collision free only for a future window $W$. The agents traverse half of this window and then redetermine their paths, this time with a circularly shifted priority order. This process goes on till all agents reach their goals. It can be noted that each agent gets the first priority periodically. It was shown in \cite{silver2005cooperative}, that WHCA* works quite well in practice in comparison to HCA*, finding solutions to almost 98\% of instances on a random map with 20\% percent obstacle occupancy. Since HCA* is the underlying method in WHCA*, a parallelizable variant of HCA* can only aid WHCA*. For that reason, simulations are limited only to HCA*. 

\section{The Problem Statement}
\noindent {\textbf{Basic assumptions}}: Consider $N$ homogeneous agents constrained to move on a Manhattan grid of size $M\times M$ (by conventions, the origin is assumed to be the top-left corner) with obstacles (an example is shown in Fig. \ref{fig:berlin_map}). Each agent has assigned source and goal locations; source location of agent $i$ is $S_i$ and its goal location is $G_i$. It is assumed that the edge traversal time is constant (say 1 unit of time) irrespective of the edge. It is also assumed that agents turn instantaneously, and are allowed to wait for an integral multiple of 1 unit of time at any of the grid points. It is further assumed that the agents stay at their goal locations indefinitely. With these, the aim is to find non-colliding paths for all agents so that the sum of the paths lengths, inclusive of waiting times, is minimum. The other cost function that is also widely used is the makespan; the maximum of the path lengths (including waiting times) across agents. For clarity, two agents collide if they end up in the same location at the same time, or if they cross each other while traversing an edge. \\
\noindent \textbf{Primer on HCA*}: The HCA* method, also known as prioritized path planning, was developed by David Silver \cite{silver2005cooperative}. As the name suggests, this method requires a priority order as a user input. As soon as the path for the first agent is determined, its positions are added to a reservation table. The second agent now finds its shortest path, ensuring that it does not collide with the first one, using the reservation table and update the same. The process continues till all agents find their respective paths. Finding the shortest path with dynamic obstacles given by the reservation table is called space-time A* and it essentially boils down to performing A* on a 3D graph. To overcome the computational complexity posed by the 3D graph, HCA* proposed the idea of Reverse Resumable A*. In this, the heuristic distance between two points used for space-time A* is the shortest distance between them on the map with static obstacles only; assuming no other agents exist. To further reduce calculations, an A* search is initiated from the goal towards the point of interest, and the present state of the nodes are stored in a dictionary. This dictionary helps in aiding the search; if the node of interest is already closed (as per A*), no calculation is done. If not, the A* algorithm resumes from where it left off, till it reaches the node of interest. As mentioned earlier, HCA* is an incomplete algorithm. That is, HCA* might not be able to find a solution under any priority order, even when a solution exists. Or, it might so happen that HCA* produces a solution for a few priority orders, and does not for others \cite{silver2005cooperative}. For this reason, the proposed method is compared with HCA* in problem instances where a solution exists for any priority order. \\
\begin{figure*}[t]
\centering
\fbox{\begin{minipage}{7in}
\begin{enumerate}
    \item Load \textit{map} onto each agent; load its \textit{source} and \textit{goal} locations; load its assigned \textit{map partition}. 
    \item Initialize the \textit{reservation table} (RT) to an empty set; initialize the set of completed agents $l$ to an empty set.
    \item {
    While $l \neq \{1,\cdots, N\}$:\\
    \hspace*{1cm}Each agent $\notin l$.\\
    \hspace*{2cm} computes and stores its \textit{shortest path} using \textit{STA*}.\\
    \hspace*{2cm} communicates the \textit{encoded shortest path} to the central server.\\
    \hspace*{2cm} computes \textit{partitions} of its shortest path.\\
    \hspace*{2cm} communicates the partitions to appropriate counterparts.\\
    \hspace*{2cm} computes the \textit{intersections of path partitions} and communicates this to the server.\\
    \hspace*{1cm} The server computes \textit{IG} from the collated paths intersection information.\\
    \hspace*{1cm} The server computes an approximation to the \textit{maximum independent set} $I$.\\
    \hspace*{1cm} The server updates $l = l\bigcup I$.\\
    \hspace*{1cm} The server updates $RT = RT\bigcup P_i$, for each $i\in I$.\\
    \hspace*{1cm} The server communicates $RT$ to each agent $\notin l$.
    }
    \item The paths \textit{stored} on each agent is its assigned path.
\end{enumerate}
\end{minipage}}
\caption{\small A pseudo-code of the proposed variant of HCA* in this paper.}
\label{main_algo}
\vspace{-0.5cm}
\end{figure*}
\noindent \textbf{Problem instance generation}: The generation of such problem instances involves several iterations; each iteration results in a source and goal location for an agent. For this purpose, let free space be defined as the set of locations which can be possible candidates for source and goal locations. Initially, the free space is the whole map except for the static obstacles. Choose a pair of source and goal locations (distinct) randomly from the free space and check if there is a path between them. If a path exists, assign the source and goal locations to the first agent; else repeat the process until a valid pair is found. Remove the path locations from the free space and update the map with the source and goal locations marked as obstacles. For the second agent, again choose a pair of source and goal locations randomly from the free space and check if a path exists between the two. If so, assign it to the second agent; else repeat till a valid pair is found. This is done for all agents. For such a problem instance, HCA* will return a solution given any priority order. This can be reasoned as follows. By construction, the source and goal of an agent $i$ does not obstruct the reachability of agent $j$ to its goal, for any $i,j$ ($i\neq j$). So, given any priority order, the agents can complete their paths one after the other. The algorithm mentioned above is outlined in Fig. \ref{instanvce_gen}.

\section{The Proposed Variant}
\begin{figure*}
    \centering
    \includegraphics[width=6in]{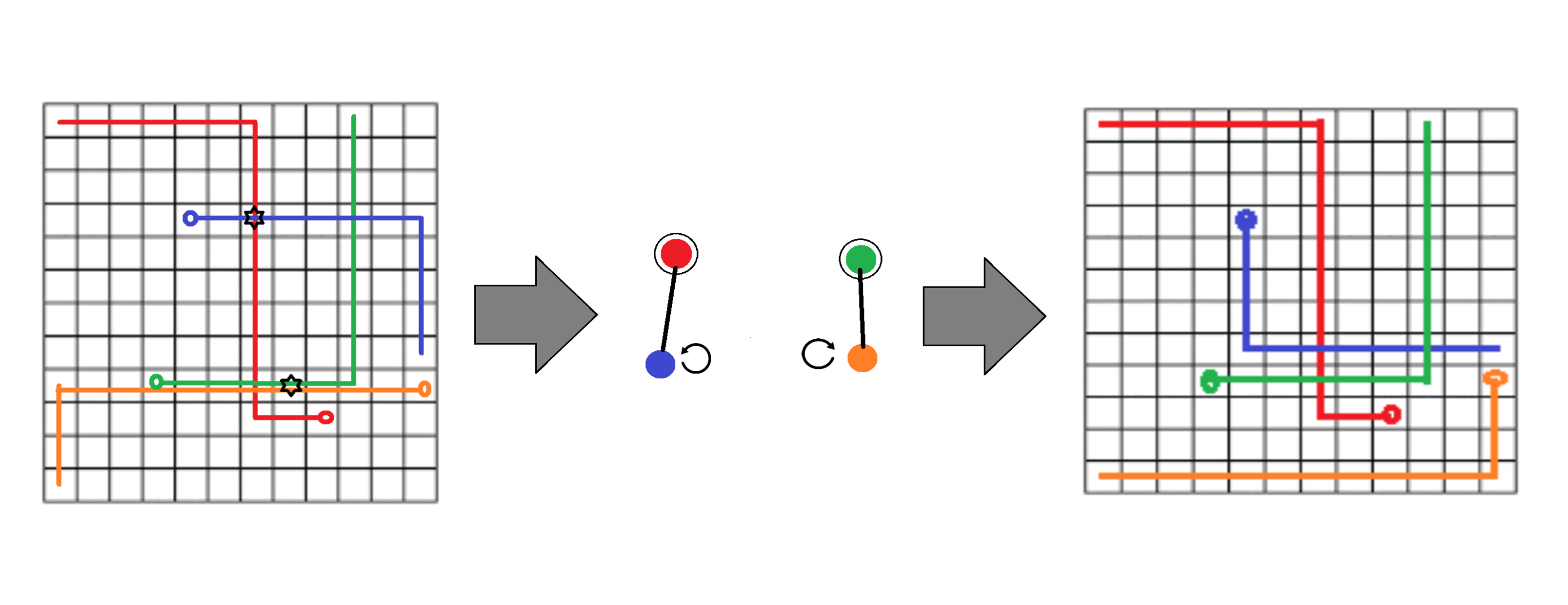}
    \vspace{-1cm}
    \caption{\small An example showing the working of the proposed method for an instance with 4 agents on $12\times 12$ grid. The grid has no obstacles for the sake of simplicity. There are four agents marked using four different colors. The starting nodes are the ones where the paths begin, and the ending nodes are the ends with small circles. It can be seen that there are two collisions in the first step, shown by stars. The IG is shown, from which an independent set is extracted (shown by encircled red and green nodes). The agents corresponding to the other nodes (blue and orange) are recomputed. The final result in shown in the image on the RHS.}
\label{working}
\vspace{-0.5cm}
\end{figure*}

\noindent \textbf{Outline of the proposed idea}: The proposed heuristic is outlined in Fig. \ref{main_algo} and an example of its working is shown in Fig. \ref{working}. It proceeds in several iterations. In the first iteration, all agents find their shortest paths disregarding the presence of others. This can be done using A*. An intersection graph (IG) is now defined from these paths in the following way. Each node in the graph represents an agent, and an edge between nodes $i$ and $j$ represent collision between the shortest paths of the $i^{\rm th}$ and the $j^{\rm th}$ agents. It is clear that an independent set from this graph would constitute a set of non-colliding paths. And that a maximum independent set would be the best, albeit greedy, choice to start with. Since finding the maximum independent set is a NP-hard problem in itself, an approximation algorithm can be used. The paths for the agents which form the independent set are fixed and the reservation table is updated. In the second iteration, the shortest paths for the remaining agents is determined using Reverse Resumable A*. By construction, these paths do not collide with those of the agents  whose paths were fixed in the first iteration. Again, an IG is constructed with these paths and an independent set is found with the approximation algorithm. As before, the paths for the agents constituting the independent set are fixed. These iterations go on till the paths for all agents are fixed. Note that in each iteration, the path of at least one agent gets fixed since an independent set has at least one node. Hence the proposed  heuristic would complete in a finite number of iterations.

\noindent \textbf{Scope for parallel compute}: Note that the path finding operations for agents in each iteration can be done in parallel. This aspect has the potential to significantly reduce computation time, which is main contribution of this work. The individual path calculations can themselves be done on the bots, since modern day bots have sufficient compute power. This shall be the assumed setup in the discussion to follow. However, this setup will also need communication between the bots and a central server. The communication latency, as a result of this, shall also be accounted for. But, if one chooses to use a multi-core processor (such as GPUs) at the central server, the bots can be spared of the computation task and the time for communication may be reduced considerably. 

\begin{figure}
    \centering
    \includegraphics[width=3.4in]{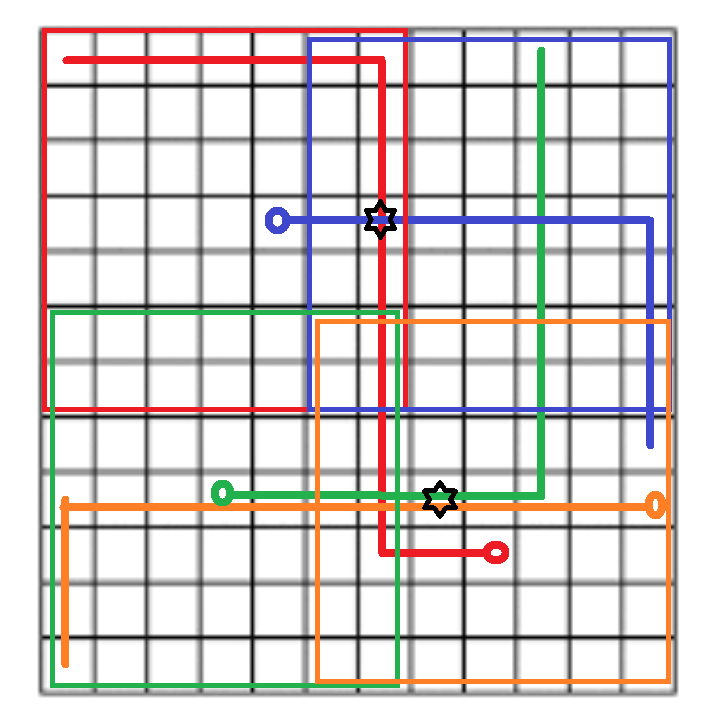}
    \vspace{-0.25cm}
    \caption{\small A schematic showing the four partitions of the map, each colored with a different color. Here each color represents the agent to which the partition is allocated. Note that the path partitions can be found out by determining the segments of paths that lie within each partition.}
    \label{fig:partition}
    \vspace{-0.5cm}
\end{figure}

\noindent \textbf{Constructing intersection graph}: The task of constructing IG from the paths can be computationally heavy, needing $^NC_2$ comparisons. However, the computation time can be reduced significantly by bringing in parallelism. Consider partitioning the map into smaller submaps. Let each path also be partitioned into subpaths; each subpath falling into a single submap. The number of subpaths falling inside a submap would typically be much smaller than the total number of paths. In addition, the lengths of subpaths would typically be much smaller as compared to the paths. These two aspects reduce the computations that need to be done to find the intersections of subpaths within a submap. If each agent is tasked with finding the subpath collisions within a particular submap, the task of finding collisions between paths can be parallelized. An example is shown in Fig. \ref{fig:partition}.

\noindent{\textbf{Partitioning the map}}: The process of partitioning the map (size of which is $b\times l$) is defined as follows. Suppose $N$ is a prime number and $l$ is the length of the longer side of the map. Then the map is partitioned into $N$ rectangular strips given by:
\begin{align}
\left\{x, y\bigg | \left[k\frac{l}{N}\right] \leq x \leq \left[(k+1)\frac{l}{N}\right], 0\leq y\leq b\right\},
\end{align}
where $0\leq k\leq N-1$.
If $N$ is not prime, then let $N = pq$ be a factorization such that $(p-\sqrt{N})^2 + (q-\sqrt{N})^2$ is minimum and $p\leq q$. For small $N$ ($\leq$ 10000, say), this factorization can be determined by a brute force search. Then the partitions are rectangular regions given by:
\begin{align} \small
    &\left\{x, y \bigg| \left[k\frac{l}{q}\right] \leq x \leq \left[(k+1)\frac{l}{q}\right], \left[m\frac{b}{p}\right] \leq y \leq \left[(m+1)\frac{b}{p}\right] \right\},
\end{align}
where $0\leq k\leq q-1$ and $0\leq m\leq p-1$. It is also noteworthy that such a partition also leads to a linear time algorithm for partitioning a path (linear in path length). That is, given $x,y$, the partition in which it lies is given by $\displaystyle \left[x\frac{N}{l}\right]$ when $N$ is prime, and is identified by the integers $\displaystyle \left[x\frac{q}{l}\right]$ and $\displaystyle \left[y\frac{p}{b}\right]$, when $N$ is not prime. 

\noindent \textbf{Finding maximum independent set}: The problem of finding the maximum independent set can be tackled in two possible ways. One way is to use approximation algorithms such as \cite{boppana1992approximating} on each connected component of the IG. Second, If each connected component is small in size ($\leq 10$), then it is also possible to run an integer linear program embodying the maximum independent set problem to determine the exact solution \cite{schrijver1998theory}. The calculation of independent set can also benefit from some level of parallelization, as  connected components can be analysed independent of each other. 

\begin{figure*}[t]
    \centering
    \subfigure[]{\includegraphics[width=0.23\textwidth]{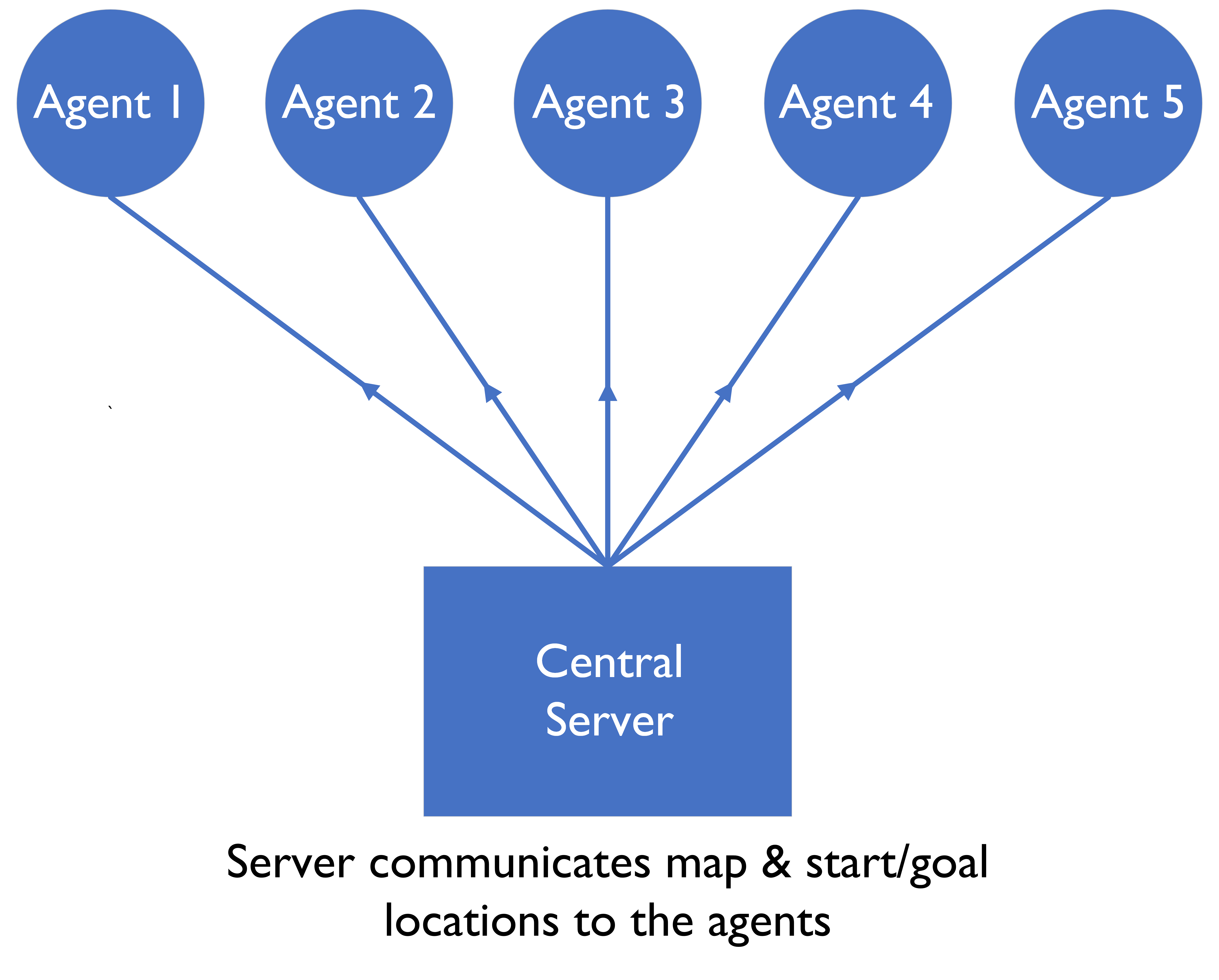}} 
    \subfigure[]{\includegraphics[width=0.23\textwidth]{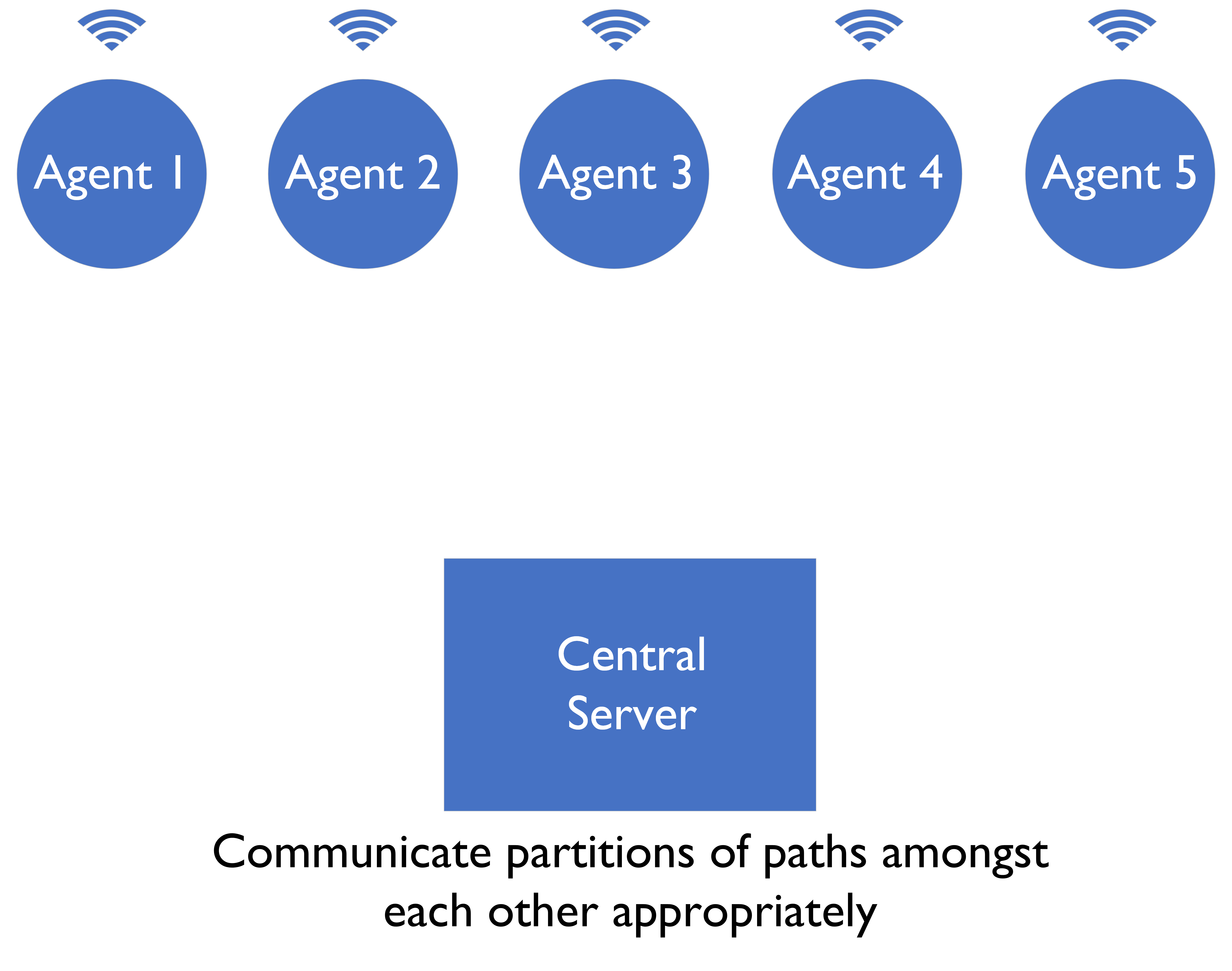}} 
    \subfigure[]{\includegraphics[width=0.23\textwidth]{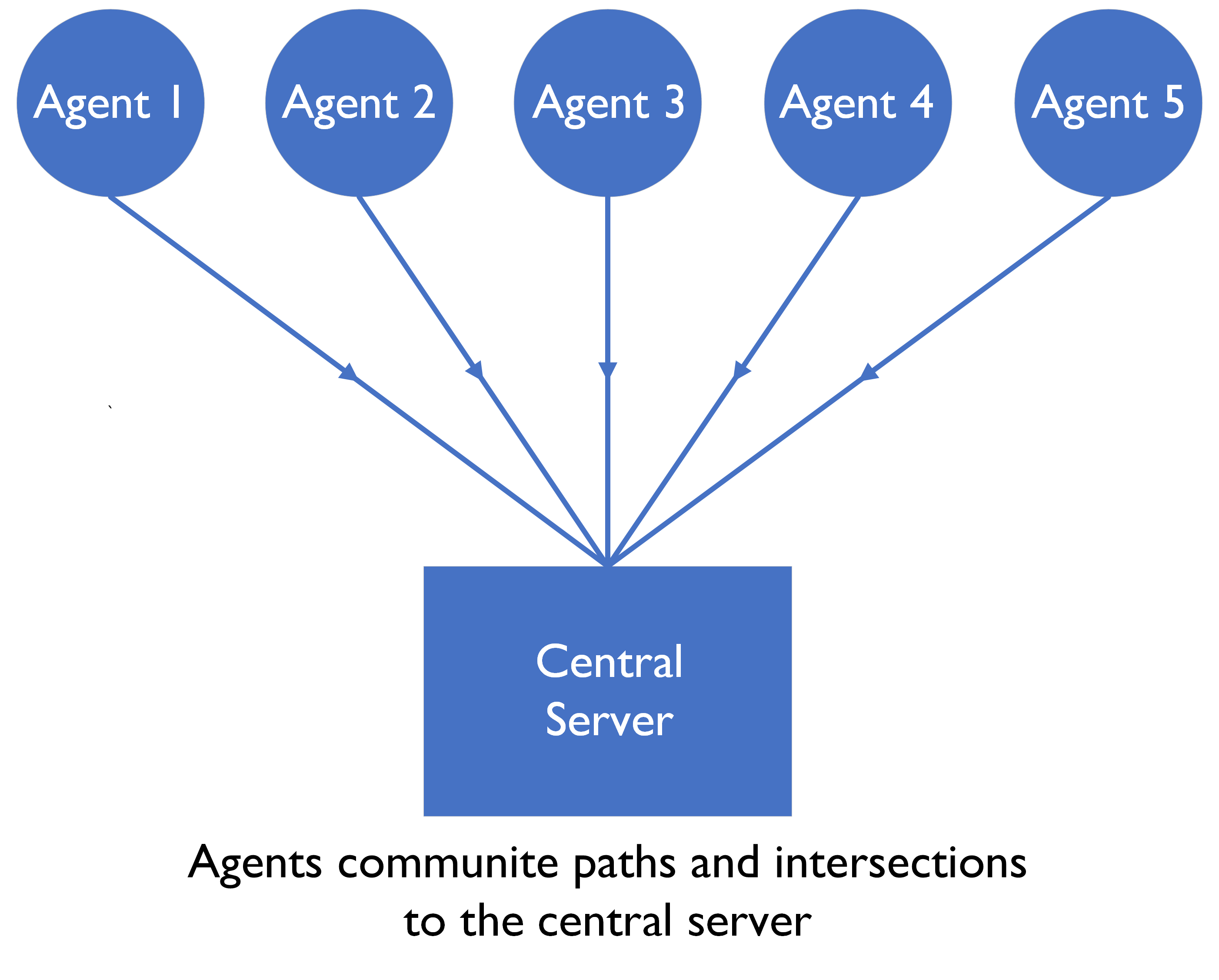}}
    \subfigure[]{\includegraphics[width=0.23\textwidth]{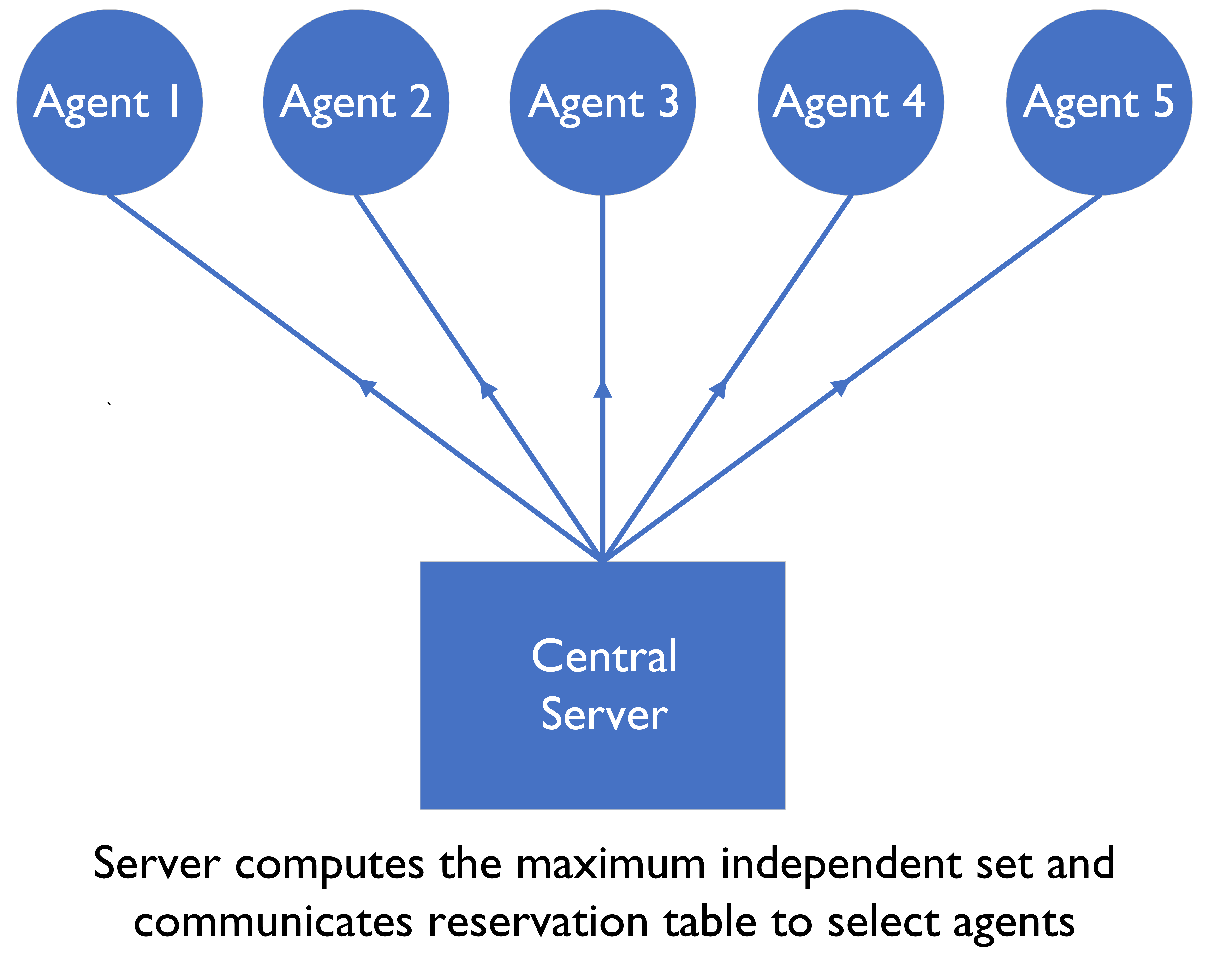}}
    \caption{\small A schematic of the working of the algorithm. In the first step shown as (a), the server communicates the map, its partitions / allocation of partitions and the start/goal locations to the agents. In the second step shown as (b), the agents compute their shortest paths, its partitions and communicate the partitions to the appropriate counterparts. In the third step shown as (c), the agents compute the intersections of path partitions and relay the information along with their respective paths to the server. In the fourth step shown as (d), the server computes an approximation to the maximum independent set, relays the update to the reservation to the agents which need to recompute their paths. This continues till all agents have found non-colliding paths.}
    \label{fig:communications}
\end{figure*}

\noindent \textbf{Encoding paths}: As mentioned earlier, the proposed method requires communication between agents and a central server. This requires that the path information be encoded efficiently, and to that end, the following encoding is proposed. There might be more efficient ways of encoding, but that is not the focus of this paper. However, it will be shown that the proposed encoding does well in practice. Consider the subpath of the $5^{\rm th}$ agent:
\begin{align}\nonumber
&(0, 0, 2), (1, 0, 3), (1, 1, 4), (1, 1, 5), (2,1,6), (2,0,7), \\\nonumber
&(3,0,8), (3, 1,9), (3, 2, 10), (3,3,11), (2,3,12),    
\end{align}
where the first two elements in a tuple represent the x-y location and the last element represents the time at which the agent was at that location. This can be encoded in the following way:
$$
5~~ 0 ~~0 ~~n~~ n~~ r~~ u~~ w~~ r~~ d~~ r~~ u~~ u~~ u~~ l~~ e,
$$
where the first integer 5 represents the agent id, the  integers 0 \& 0 represent the x \& y coordinates of the starting location of the subpath, the number of  $n$'s that follow thereafter represent the starting time of the subpath (2 in this case), $r$ represents an increase in the x-coordinate (or, moving to the right), $l$ represents a decrease in the x-coordinate (or, moving to the left), $u$ represents an increase in the y-coordinate (or, moving up), $d$ represents a decrease in the y-coordinate (or, moving down), $w$ represents a wait and finally, $e$ represents the end of subpath segment. It can be argued that the proposed encoding would require lesser bits as compared to that required for  raw data. A path of length $L$ would require $(3L + \delta)$ bits, where $\delta$ is the number of bits required to encode the agent number, starting location, the starting time and the end of path. However, for raw data, one would require $(2\log_2(M)L + \delta)$. For a large $M$, the former would be much smaller than the latter. If a subpath inside a submap comprises of more than one time-contiguous segments, each of these segments is encoded separately. \\
\noindent \textbf{Communication cost}: The communication cost here is defined as the total number of bits that need to be transmitted between the agents and the central server during the execution of the heuristic. The sequence of communications is outlined in Fig. \ref{fig:communications}. It is assumed that the map and its partitions are loaded on all agents just once, and its communication cost is insignificant over time, compared to the communication cost incurred in repeated path finding computations. Firstly, in every iteration, the number of bits needed to communicate the source and goal locations is
$2N\log_2(M)$. 
This information is communicated by the central server to the agents, one after the other. Secondly, the agents need to communicate the path partitions to each other (appropriately, for the purpose of intersection detection) and also to the central server (for the purpose updating reservation table). For this, the agents take turns in a pre-defined order to communicate the information. To calculate the bits transferred in the second step, the number of bits needed to encode paths must be calculated. Note that the paths are encoded in the way that was described earlier. 
Suppose in the $r^{\rm th}$ iteration, the path of the $i^{\rm th}$ agent has $s^r_i$ time-contiguous subpath segments (in order), each with a length of $l^r_{k,i}$ ($1\leq k\leq s^r_i$). The number of bits needed to encode the $k^{\rm th}$ segment of the path for the $i^{\rm th}$ agent in this iteration would be at most
\begin{align}
\lceil\log_2(N)\rceil + 2\lceil\log_2(M)\rceil + 3\sum_{j=1}^{k-1}l^r_{j,i} + 3(l_k+1).
\end{align}
The first term is the number of bits needed to encode the agent number, the second term is the number of bits needed to encode the initial location, the third term is the number of bits needed to encode \textit{n}'s and the last term is the number of bits needed to encode the directions (and the end of path). 
Thus, the number of bits needed to encode the full path for agent $i$ in the $r^{\rm 
 th}$ iteration would be
\begin{align}
b^r_{i} = \sum_{k=1}^{s^r_i}  \left(\lceil\log_2(N)\rceil + 2\lceil\log_2(M)\rceil + 3\sum_{j=1}^{k-1}l^r_{j,i} + 3(l_k+1)\right).
\end{align}
And thus again, the total number of bits communicated as part of the second step in the $r^{\rm 
 th}$ iteration is 
\begin{align}
\sum_{i=1}^{N}\delta_{i,r} b^r _{i},
\end{align}
where $\delta_{r,i}$ is one if agent $i$'s path is not yet fixed in the $r^{\rm th}$ iteration, and zero otherwise.
Thirdly, there is the communication of the path intersections from the agents to the central server. Suppose, in the $r^{\rm th}$ iteration, agent $i$ finds $E^r_i$ intersections.
Then, the number of bits needed to communicate this information (all agents put together) is 
\begin{align}
b^r_{IG} = \sum_{i=1}^N 2\lceil\log_2(N)\rceil E^r_i.
\end{align}
Lastly, in each iteration the central server also needs to communicate the reservation table (RT) to all the agents pending path assignment. Since the same RT is communicated to all these agents, the information can be broad-cast by the server. And since the reservation table consists of all fixed paths it should clear that the total number of bits communicated, in all iterations put together, would be at most the sum of bits needed to encode all fixed paths. That is, if the length of the fixed path of the agent $i$ is $L_i$, then the communication cost for transferring the RT would be
\begin{align}
b_{RT} = \sum_{i=1}^N \lceil\log_2(N)\rceil + 2\lceil\log_2(M)\rceil + 3(L_i+1).
\end{align}
And if the total number of iterations is $R$ and if the data rate is $d$, the time taken for communication would be
\begin{align}
C_{comm} = \frac{1}{d}\left(b_{RT} + \sum_{r=1}^{R}\left( b^r_{IG} +  \sum_{i=1}^{N}\delta_{r,i}b^r_{i}  \right) \right).
\end{align}
Needless to say, the speedup from parallelism depends crucially on the implementation of the algorithm and the data rate for communication. If the computation time of HCA* is $C_0$, the computation time of the proposed algorithm is $C$, then the speedup is given by:
\begin{align}
\frac{C_0}{C + C_{comm}}.
\end{align}
In other words, the proposed heuristic might not outperform HCA* if the data rates are low.  

\begin{table*}
\begin{center}
\resizebox{7in}{!}{\begin{tabular}{|c|c|c|c|c|c|c|}
\hline
Map Type & Map Size & \# of Agents & \# of Expts & Sum of Costs Ratio & Makespan Cost Ratio & Compute Time Ratio \\
& & & & (Avg, Min, Max, Median) & (Avg, Min, Max, Median) & (Avg, Min, Max, Median) \\
  \hline
 Random (0.1) & 100*100 & 64 & 100 & (0.9988, 0.9930, 1.0015, 0.9993) & (0.9999, 0.9925, 1.0061, 1.0) & (0.2812, 0.0805, 1.3673, 0.2587) \\  
 \hline
Random (0.2) & 100*100 &  64 & 100 & (0.9983, 0.9916, 1.0012, 0.9988) & (1.0001, 0.9862, 1.0150, 1.0) & (0.3107, 0.0729, 0.8731, 0.3019) \\ 
 \hline
Berlin & 100*100 & 64 & 100 & (0.9963, 0.9827, 1.0026, 0.9973) & (1.0, 0.9810, 1.0270, 1.0) & (0.3391, 0.0055, 11.8839, 0.1801) \\  
 \hline
Warehouse & 161*63 & 64 & 100 & (0.9734, 0.9277, 1.0115, 0.9744) & (0.9982, 0.9009, 1.0659, 1.0) & (0.6829, 0.0022, 28.41, 0.0616) \\ 
 \hline
\end{tabular}}
\end{center}
\label{tab1}
\caption{\small \small Table showing comparative results between the proposed algorithm and HCA*. The first columns shows the map type \cite{stern2019mapf}, the second column shows the size of the map, the third shows the number of agents, the fourth shows the number of problem instances run, the fifth column shows the statistics of ratio of the sum of costs of the proposed algorithm to that obtained from HCA*, the sixth shows those for makespan and the last column shows those for the total compute time.}
\vspace{-0.5cm}
\end{table*}

\section{Simulations}
\noindent \textbf{Simulation setup}: The simulation results are presented in Table I. For these, the package published as part of \cite{antoniazzi2020implementation} was used. The simulation setup is described as follows. Four different maps are chosen for bench marking the proposed heuristic against HCA*: (i) random map generated with a probability of obstacle equal to 0.1, (ii) random map generated with a probability of obstacle equal to 0.2, (iii) the Berlin map, and (iv) the warehouse map. See \cite{stern2019mapf} for more details. The Berlin map and the warehouse map were down-sampled from the original size of $256\times 256$ to $100\times 100$ for ease of compute. For each map, several different problem instances were generated as per the method outlined previously, and results were collected for 100 instances where both the proposed heuristic and HCA* found paths within a stipulated time. Also, for each run in the random maps cases, a different map was generated. The number of agents was chosen to be 64, and therefore the map partitions were computed by dividing each side of the map into 8 parts. For HCA*, a random priority order was chosen for each run. Finally, it is assumed that the agents can communicate at the rate of 10 MBps.   

\noindent \textbf{The Results}: For each map, the average, the minimum, the maximum and the median of the following are shown: (i) the ratio of sum of costs obtained with the proposed heuristic to that obtained from HCA*, (ii) the ratio of makespan costs obtained with the proposed heuristic to that obtained from HCA*, (iii) the ratio of the compute time for the proposed heuristic to that of HCA*. It can be noted that the average and median ratios for cost functions are less than or equal to 1, implying that the proposed heuristic does better than HCA* in several instances. An empirical comparison with CBS might be warranted if the reader wishes to compare the obtained cost with the optimal value, but this exercise is out of the scope of this paper, due to its computational complexity. As far the compute time is concerned, the average and median ratios suggest a speedup of upto 4x, approximately.  



\section{Conclusions}
In this paper, a parallelizable variant of HCA* was proposed. The central idea was to let agents calculate shortest paths independently in several iterations, and then let a central server decide on which ones to fix in each iteration. This aspect created a scope for parallel compute. The central server decided on the paths to be fixed by solving a maximum independent set problem on an intersection graph. This aspect also frees the user of defining a priority order. The intersection graph, in turn, was also calculated in parallel by the agents. The act of dividing the map into several partitions made this possible. The agents were assumed to be equipped to communicate with each other and the central server. The communication latency was also accounted for in the simulations. The simulations showed that the proposed heuristic outperforms HCA* (cost wise) in several instances  and can provide a significant speedup in execution. It might also be interesting to use the proposed heuristic for WHCA* with a change. That is, determine the first iteration maximum independent set with a particular agent being part of it. 
\bibliographystyle{IEEEtran}
\bibliography{references}


\end{document}